\documentclass[preprint,prb,showpacs,superscriptaddress]{revtex4-1}
\usepackage{amsmath}
\usepackage{graphicx}
\usepackage{amssymb}
\usepackage{times}
\usepackage{color}
\usepackage{multirow}
\definecolor{lr}{rgb}{1.0,0.3,0.3}
\definecolor{dg}{rgb}{0.0,0.5,0.0}

\usepackage{comment}

\graphicspath{./}
 


\begin{document}

\title{Hybrid-DFT+V$_w$ method for accurate band structure of correlated transition metal compounds: the case of cerium dioxide}

\author{Viktor Iv\'ady}
\email{vikiv@ifm.liu.se}
\affiliation{Department of Physics, Chemistry and Biology, Link\"oping
  University, SE-581 83 Link\"oping, Sweden}
\affiliation{Wigner Research Centre for Physics, Hungarian Academy of Sciences,
  PO Box 49, H-1525, Budapest, Hungary}

\author{Adam Gali} 
\affiliation{Wigner Research Centre for Physics, Hungarian Academy of Sciences,
  PO Box 49, H-1525, Budapest, Hungary}
\affiliation{Department of Atomic Physics, Budapest University of
  Technology and Economics, Budafoki \'ut 8., H-1111 Budapest,
  Hungary}

\author{Igor A. Abrikosov}
\affiliation{Department of Physics, Chemistry and Biology, Link\"oping
  University, SE-581 83 Link\"oping, Sweden}
\affiliation{Materials Modeling and Development Laboratory, National University of Science and Technology `MISIS', 119049 Moscow, Russia}

\date{\today}

\begin{abstract}
Hybrid functionals' non-local exchange-correlation potential contains a derivative discontinuity that improves on standard semi-local density functional theory (DFT) band gaps. Moreover, by careful parameterization, hybrid functionals can provide self-interaction reduced description of selected states. On the other hand, the uniform description of all the electronic states of a given system is a know drawback of these functionals that causes varying accuracy in the description of states with different degrees of localization. This limitation can be remedied by the orbital dependent exact exchange extension of hybrid functionals; the hybrid-DFT+V$_w$ method [V. Iv{\'a}dy, et al., Phys. Rev. B 90, 035146 (2014)]. Based on the analogy of quasi-particle equations and hybrid-DFT single particle equations, here we demonstrate that parameters of hybrid-DFT+V$_w$ functional can be determined from approximate quasi-particle spectra. The proposed technique leads to a reduction of self-interaction and provides improved description for both $s$ / $p$ and $d$ / $f$-electrons of the simulated system. The performance of our charge self-consistent method is illustrated on the electronic structure calculation of cerium dioxide where good agreement with both quasi-particle and experimental spectra is achieved.
\end{abstract}
\maketitle


\section{Introduction}

Density functional theory (DFT) with approximate exchange-correlation functionals has proven to be a good compromise between predictive power and computational efficiency\cite{KohnNobelL}. On the other hand, to further improve accuracy, one must solve longstanding issues of DFT, such as the  missing derivative discontinuity of local and semi-local exchange-correlation functionals\citep{PerdewLevy1983,Sham1983,Perdew91} or self-interaction of the Kohn-Sham particles\citep{PerdewZunger1981}. The latter error appears due to non-zero sum of the self-repulsion and self-attraction due the Hartree and exchange-correlation interactions, respectively. Manifestations of the self-interaction are, for example, the over delocalization of the states, the spurious curvature of the total energy curve when generalized to non-integer occupation numbers and the violation of the generalized Koopmans' theorem\citep{LanyZunger09,LanyZunger10,Dabo}, in other context, the ionization potential theorem\citep{Perdew91, Perdew97,Almbladh85}. To overcome DFT's self-interaction problem, various methods have been proposed, for instance, self-interaction corrected functionals\citep{PerdewZunger1981,SvaneSIC90} and Koopmans-compliant functionals\cite{KCfunctionals}.

Hybrid functionals that mix local or semi-local DFT exchange potential with the non-local exact exchange potential of single particles\cite{Becke1993} have became popular in several fields of condensed matter physics and quantum chemistry\cite{Kummel08}. The mixing of the potentials is governed by adjustable parameters that are standardly determined by theoretical considerations or fitting to experimental data sets.  From the theoretical point of view, hybrid functionals' success can be accounted for by the introduction of derivative discontinuity in the exchange-correlation potential and the partial or complete reduction of self-interaction error\cite{Kummel08}. Recently, it was shown that by tuning hybrid functionals' internal parameters, self-interaction free description of selected single particle states is possible\cite{Kronik2012,KronikPRL12,Deak2017}. Furthermore,  the strategy to determine hybrid functionals mixing parameters from higher level self-interaction free \emph{ab initio}  theory is frequently applied nowadays. \cite{Shimazaki2009,Marques2011,Kronik2012,KronikPRL12,Atalla13,Refaely2015,BrawandPRX2016}

Hybrid functionals, on the other hand, are global functionals in the sense that all the electronic states are described uniformly, i.e.\  by using the same functional parameters.\cite{IvadyPRB2013} When states of different degrees of localization,  such as $sp$-hybridized on one hand and $d$ and $f$-orbital related states on the other hand are simultaneously present in the system, uniform treatment of all the states is an inaccurate approximation. We emphasize that such system cannot be described accurately simply by tuning the mixing parameters of hybrid functionals. These issues arise, for example, for transition metals and rear earth oxides and $sp$ hybridized systems, like conventional semiconductors or organic molecules, that include transition metal impurities. To overcome these limitations, one must go beyond the traditional global hybrid approximation\cite{Perdew07}, for example by locally changing \cite{Jaramillo03,Krukau08} the mixing of the exact and (semi-)local exchange of the generalized Kohn-Sham particles.

As an alternative approach, hybrid-DFT+V$_w$ method was recently introduced through the unification LDA+$U$ and hybrid-DFT methods.\cite{IvadyPRB2014} Hybrid-DFT+V$_w$ method can be considered as orbital dependent extension of hybrid-DFT, similarly to the case of LDA and LDA+U methods. It employs hybrid exchange-correlation description for the couplings of delocalized states, while the couplings between the localized states is corrected by the $V_w$ potential. The functional internal parameter $w$ must be determined for the considered system. One possible way to obtain $w$ and to make the functional self-interaction corrected is to enforce the fulfillment of the generalized Koopmans' theorem (gKT)\cite{Dabo,IvadyPRB2013}.  However, checking the fulfillment of the gKT requires the calculation of charged systems that unavoidably introduces finite size effects both in the total and the generalized Kohn-Sham energies when periodic boundary condition is applied. Furthermore, the Koopmans' theorem is well-defined for localized states rather than for extended states\citep{Almbladh85}. Thus additional problems arise when the functional is applied to bulk systems. Relying on gKT, there are only limited possibilities to determine the parameter of hybrid-DFT+V$_w$ method, thus it limits the its applicability.

Here, we propose an alternative strategy to determine $w$ and to make hybrid-DFT+V$_w$ functional self-interaction corrected. We discuss the analogy between the quasi-particle equations and the hybrid-DFT single particle equations that allows us to utilize higher level self-interaction free theory, such as the Coulomb hole and screened exchange (COHSEX)  and G$_0$W$_0$ approximation\cite{Hedin} of the many-body perturbation theory (MBPT), to determine the $w$ parameter. The provided approach does not rely on the calculation of charged systems, thus it can be applied to both extended and localized systems to obtain self-interaction corrected charge self-consistent solutions with affordable computational cost.

As an example we calculate the electronic structure of cerium dioxide (CeO$_2$). Cerium dioxide includes $p$, $d$ and $f$-electrons close to the Fermi energy\cite{Andersson2007, Scheffler09}, thus it is an ideal model system for our method. Furthermore, understanding cerium oxides electronic structure, including CeO$_2$ and Ce$_2$O$_3$, is particularly important as these materials are frequently used, for example in automotive exhaust treatment.\cite{Campbell713} Creation and annihilation of oxygen vacancies play important role in the oxidation processes\cite{Skorodumova02,Campbell713}, thus an accurate characterization of cerium atoms oxidation state and the energetics of oxygen vacancy in CeO$_2$ is highly desirable. Determination of these parameters requires  computationally efficient charge self-consistent first principles methods that can describe both defects and bulk states accurately. We demonstrate here that hybrid-DFT+V$_w$ method, which has already proven its accuracy for point defect calculations in conventional semiconductors\cite{IvadyPRB2013,IvadyPRB2014}, is a appropriate tool for such investigations.

The article is organized as follows: In section II we discuss the theoretical background and introduce our new approach. This section includes the discussion of hybrid-DFT single particle and MBPT quasi-particle equations similarity, a short review of the hybrid-DFT+V$_{w}$ method, and the introduction of first principles  determination of the $w$ parameter. In section III  and section IV we describe the details of our computational methods and present our results, respectively. Finally, in section V we summarize our findings.

\section{Theory}  
  
 \subsection{Analogy of generalized Kohn-Sham particles and MBPT quasi-particles} 
  
Although hybrid functionals have firm theoretical foundations though the generalized Kohn-Sham schemes\cite{GKS1996}, at the birth of hybrids there were no good theoretical recipes for the determination of the mixing of the exact and local exchange potentials, thus hybrid functionals were often constructed with free parameters that were usually determined by fitting to experimental data sets.\cite{Kummel08} 
Let us consider, for example, the form of the PBE0\cite{PBE0}  hybrid functional potential: 
\begin{equation} \label{eq:pbe0_pot}
V_{xc}^{\text{PBE0}}\! \left(\mathbf{r}, \mathbf{r}' \right) = \alpha V_{x}^{\text{exact}}\! \left(\mathbf{r}, \mathbf{r}' \right) + (1-\alpha) \delta\!\left(\mathbf{r} - \mathbf{r}'\right) V_{x}^{\text{DFT}} \! \left( \mathbf{r}\right) + \delta\!\left(\mathbf{r} - \mathbf{r}'\right) V_{c}^{\text{DFT}} \left( \mathbf{r}\right) \text{,}
\end{equation}
where $V_{x}^{\text{exact}}$, $V_{x}^{\text{DFT}}$, and $V_{c}^{\text{DFT}}$ are the non-local exact exchange and the local DFT exchange and correlation potentials, respectively, $\delta\!\left(\mathbf{r} - \mathbf{r}'\right) $ is the Dirac delta function, and $\alpha $ is the mixing parameter. It is clear that only fraction of the exact exchange interaction, or in other words, exact exchange with a screen Coulomb potential is introduced. To understand why this can work, one can consider an analogy between the generalized Kohn-Sham particles and the MBPT quasi-particles. Despite the lack of direct theoretical connection, the equations have similar form thus an analogy can be made. 

The MBPT quasi-particle equation can be written as
\begin{equation}
\hat{H}_{0} \psi_i \! \left( \mathbf{r} \right) + \int \Sigma\!\left( \mathbf{r}, \mathbf{r}', \varepsilon_i \right) \psi_{i}\!\left(\mathbf{r}'\right)  d^{3}r' = \varepsilon_i \psi_i\!\left(\mathbf{r}\right)  \text{,}
\end{equation}
where $\psi_i \! \left( \mathbf{r} \right) $  is the quasi-particle amplitude, $\hat{H}_{0}$ is the Hamiltonian of non-interacting particles, and $\Sigma\!\left( \mathbf{r}, \mathbf{r}', \varepsilon_i \right)$ is the non-local and energy dependent quasi-particle self-energy, which can be obtained exactly by Hedin's equations\cite{Hedin}, but approximated in practice. The most frequently used approximation is the GW approximation\cite{Hedin}, $\Sigma  = i GW $, where $G$ is the Green's function and $W$ is the screened Coulomb interaction. Starting with Eq.~(\ref{eq:pbe0_pot}), the equation of the interacting generalized Kohn-Sham particles can be written in a similar form,
\begin{equation}
\hat{H}_{0} \varphi_i \! \left( \mathbf{r} \right) + \int V_{\text{xc}}^{\text{PBE0}}\!\left( \mathbf{r}, \mathbf{r}' \right) \varphi_i\!\left(\mathbf{r}'\right)  d^{3}r' = \varepsilon_i \varphi_i\!\left(\mathbf{r}\right)  \text{,}
\end{equation}  
where $\varphi_i$ are the general Kohn-Sham particles wavefunctions. 

In static or Coulomb hole and screened exchange (COHSEX) approximation\cite{Hedin} of the GW self-energy, the self-energy can be written in the form of 
\begin{equation}  \label{eq:COHSEX}
\Sigma_{\text{COHSEX}} \! \left( \mathbf{r}, \mathbf{r}' \right) = \Sigma_{\text{SEX}} \! \left( \mathbf{r}, \mathbf{r}' \right) + \Sigma_{\text{COH}} \! \left( \mathbf{r}, \mathbf{r}' \right) \text{,}
\end{equation}
with
\begin{equation}  \label{eq:SEX}
\Sigma_{\text{SEX}} \! \left( \mathbf{r}, \mathbf{r}' \right) = - \sum_{j}^{\text{occ.}} \psi_j \!\left( \mathbf{r} \right) \psi_j^{*} \!\left( \mathbf{r}' \right) W \! \left( \mathbf{r},\mathbf{r}' \right) \text{,}
\end{equation}
where $W \! \left( \mathbf{r}, \mathbf{r}'\right)$ describes the static screened interaction potential of the quasi-particles, and
\begin{equation}  \label{eq:COH}
\Sigma_{\text{COH}} \! \left( \mathbf{r}, \mathbf{r}' \right) = \frac{1}{2} \delta \! \left( \mathbf{r} - \mathbf{r}'\right) W_{p} \! \left( \mathbf{r}, \mathbf{r}' \right) \text{,}
\end{equation} 
where $W_{p} \! \left( \mathbf{r}, \mathbf{r}'\right) = W \! \left( \mathbf{r}, \mathbf{r}'\right)  -  v_{ee} \! \left( \mathbf{r} -\mathbf{r}'\right) $ and $ v_{ee} \! \left( \mathbf{r} -\mathbf{r}'\right) $ is the bare Coulomb interaction potential.

The first term on the right hand side of Eq.~(\ref{eq:COHSEX}) describes the static exchange interaction between the quasi-particles. Importantly, this terms includes a screened, non-local interaction potential $W \! \left( \mathbf{r}, \mathbf{r}'\right)$. The second term describes the interaction with the Coulomb hole, which forms around the electrons due to correlation effects. This interaction is approximated with a local potential, see Eq.~(\ref{eq:COH}). The self-energy in the COHSEX approximation shows close similarity to the hybrid functionals' mixed non-local and local exchange-correlation potential. In the PBE0 functional,  $ \alpha V^{\text{ex}}_{\text{x}} \! \left( \mathbf{r}, \mathbf{r}' \right)$ can be considered as an approximations to $\Sigma_{\text{SEX}} \! \left( \mathbf{r}, \mathbf{r}'\right)$, while the semi-local PBE exchange-correlation part given by the second and the third terms in the rhs of Eq.~(\ref{eq:pbe0_pot}) can be viewed as an analogy  to $\Sigma_{\text{COH}} \! \left( \mathbf{r}, \mathbf{r}'\right)$. Based on this analogy, the following statement can be made: The generalized Kohn-Sham particles are approximations to the quasi-particles, therefore, the generalized Kohn-Sham eigenvalues $\varepsilon_i$ can be considered as rough approximations to the quasi-particle energies.\cite{KronikPRL12,Atalla13} These arguments were utilized recently to determine the mixing parameter of hybrid functionalism from the screened exchange interaction potential or from the inverse dielectric matrix obtained by approximate MBPT calculation.\cite{Atalla13,BrawandPRX2016} On the other hand, as the hybrid functionals are not directly derived from the quasi-particle equation, the validity of this analogy can only be determined in practical applications.

 \subsection{The hybrid-DFT+V$w$ method} 

As we have seen above an analogy can be made between hybrid-DFT and the static COHSEX approximation of many-body perturbation theory. Furthermore, it is well know that LDA+U method can be also connected to the COHSEX approximation\cite{Anisimov1997,RinkeLDAU10}. Consequently, there is an indirect connection between hybrid-DFT and LDA+U methods. Moreover, a direct mathematical link between the two approaches was recently revealed in Ref.~\onlinecite{IvadyPRB2014}. Here, we shortly review the basic idea of the derivation and introduce the hybrid-DFT+V$_w$ scheme.
   
An important difference between the LDA+U and hybrid+DFT approach is that the LDA+U method acts differently on a subspace of correlated orbitals, while hybrids are global exchange-correlation functionals. To be able to make a connection, one can consider the effect of the introduction of the exact exchange potential only on the subset of localized atomic-like orbitals $\phi_{m}$, where $m$ is the projection of the orbital momentum. Furthermore, as LDA+U was introduced as a correction to a local DFT exchange-correlation functional, for the sake of comparison,  we express the mixing of the exact and approximate DFT exchange in hybrids as a correction to the DFT potential.

 By rearranging the terms in Eq.~\ref{eq:pbe0_pot} the PBE0 exchange-correlation potential can be written as:
\begin{equation}\label{eq:hybridcorr}
V_{\text{xc}}^{\text{PBE0}} \! \left( \mathbf{r}, \mathbf{r}' \right) = \alpha \left( V_{\text{x}}^{\text{ex}} \! \left( \mathbf{r}, \mathbf{r}' \right) - \delta \! \left( \mathbf{r} -\mathbf{r}' \right)  V_{\text{x}}^{\text{PBE}} \! \left( \mathbf{r} \right)  \right)  + \delta \! \left( \mathbf{r} -\mathbf{r}' \right)   V_{xc} ^{\text{DFT}}\! \left( \mathbf{r}\right)  \text{,}
\end{equation}
where the first and second terms define the potential correction to the DFT exchange-correlation potential $V_{xc}^{\text{DFT}}$.
To calculate the correction on the subset of correlated orbitals, we define the on-site occupation matrix 
\begin{equation} \label{eq:occmatrix}
\mathbf{n}_{mm'}^{\sigma} = \sum_{i} \left\langle \varphi_i \left|  \phi_m \right.  \right\rangle \left\langle  \left. \phi_{m'} \right| \varphi_i  \right\rangle \text{.}
\end{equation}
By assuming that an appropriate unitary transformation of the of the atomic orbitals ${\phi'}_{m} = \hat{U}^{\dagger}\phi_{m}$ diagonalizes the occupation matrix, one can consider only the diagonal elements $n_{m}^{\sigma}$ and rewrite the exact exchange potential as
\begin{equation}
V^{\textup{ex}}_{x} \! \left [  n^{\sigma }_{m}  \right ] = - \sum_{  m', \sigma }  \left \langle m m' \left | v_{ee} \right | {m}' m \right \rangle    n_{m'}^{\sigma } \text{.}
\end{equation}

The second term on the right hand side of Eq.~(\ref{eq:hybridcorr}) corrects for the double counting. To calculate this term, one should determine the restricted effect of the semi-local exchange-correlation functional on the subset of correlated orbitals. As this cannot be made explicitly, an approximation is employed\cite{FLL1994,Liechtenstein1995,IvadyPRB2014}, similarly to the one of the LDA+U method. However, in contract to the LDA+U method, where the screened Hubbard $U$ and Stoner $J$ parameters are used, here the bare interaction strengths, $F_0$ and $J_0$ are used to be consistent with the restriction, in which only interactions among the localized orbitals are taken into account. Thus the screening effect of other delocalized states is not considered. With these in mind, the second term on the right hand side of Eq.~(\ref{eq:hybridcorr}) can be approximated\cite{FLL1994,Liechtenstein1995,IvadyPRB2014} as
\begin{equation}
V_{\text{xc}}^{\text{PBE}} \! \left [ n^{\sigma_{m} } \right ] = - \frac{F^{0} - J^{0}}{2} - J^{0} n^{\sigma }_{m} \text{.}
\end{equation} 

By applying the approximation $\left\langle mm \left|v_{ee} \right|  mm  \right\rangle \approx F_0 $ and $\left\langle mm' \left|v_{ee} \right|  m'm  \right\rangle \approx J_0 $ \cite{IvadyPRB2014} the localized orbital-restricted form of the potential correction term in Eq.~(\ref{eq:hybridcorr}) can be written as
\begin{equation} \label{eq:lochybridcor}
\Delta \! V_{m}^{\textup{PBE0x}, \sigma} \! \left [ \mathbf{n}_{m}^{\sigma }   \right ]    =  \alpha  \left (  F^{0} -  J^{0} \right ) \left ( \frac{1}{2} - \mathbf{n}_{m}^{\sigma } \right ) \text{.}
\end{equation}

The above equations show close similarities to the potential correction terms of the LDA+U method by Dudarev \emph{et al.}\ \cite{Dudarev1998}. From the mathematically equivalent forms of the corrections, one can deduce that the two methods have the same effect on localized atomic-like orbitals. The difference is only in the strength of the correction. While in the LDA+U method, the correction strength is determined by the effective Hubbard $U$ parameter, which is usually set by hand, in hybrids the magnitude of the correction is determined by the mixing parameter $\alpha$. 

In such systems, where both $sp$-hybridized and localized $d$ or $f$-like states are present, a single $\alpha$ parameter most often is not sufficient to provide a suitable description for orbitals of different degrees of localization.  To overcome the limitations of the uniform mixing approximation of usual hybrids, the hybrid-DFT+V$_w$ method was proposed to accomplish this in an orbital dependent fashion\cite{IvadyPRB2013,IvadyPRB2014}. 
An on-site potential is added to a hybrid functional exchange-correlation potential in the form of Eq.~(\ref{eq:lochybridcor}) as
\begin{equation}
V_{w} \! \left [ \mathbf{n}_{m}^{\sigma } \right ]    =  w \left ( \frac{1}{2} - \mathbf{n}_{m}^{\sigma } \right ) \text{,}
\end{equation}
where $w$ defines now the strength of the on-site potential correction. To achieve self-interaction reduced description
\begin{equation} \label{eq:wdef}
w = U^{\text{eff}}_{\text{real}} - U^{\text{eff}}_{\text{hybrid}} = U^{\text{eff}}_{\text{real}} -  \alpha \left( F^{0} -J^{0} \right)  \text{,}
\end{equation}
must be satisfied, where $U^{\text{eff}}_\text{{real}}$ is the true effective on-site interaction strength. On the other hand, as  $U^{\text{eff}}_\text{{real}}$ is generally unknown, $w$ can also be considered as an adjustable parameter. Note that, the $V_w$ correction can be translated to an orbital dependent adjustment of the mixing parameter $\alpha$ through the formula of
\begin{equation}
\alpha_{\text{loc}} = \alpha + \frac{w}{F^{0}-J^{0}} \text{.}
\end{equation}

Alternatively, one can determine the $w$ parameter through the fulfillment of the theoretical requirements of the exact DFT functional. One of these requirements is the so-called ionization potential (IP) \cite{Perdew91,Perdew97} or, in other context, the generalized Koopmans' theorem (gKT)\cite{LanyZunger09,Dabo}. According to these theorems the highest occupied Kohn-Sham eigenvalue is equal to the negative ionization energy of the system, and it stays constant under the variation of its occupation number. When these criteria are fulfilled the functional is approximately self-interaction free, at least for the highest occupied orbital. Standard approximate exchange-correlation functionals do not satisfy gKT, however, there are purpose built functionals that do fulfill it, such as Koopmans-compliant (KC) functionals\cite{Dabo,KCfunctionals}. In such cases, when a suitable correction potential is known, by adjusting its parameters the gKT can be fulfilled to achieve self-interaction corrected description.

This strategy was applied to determine the $w$ parameter and to calculate the charge transition levels of the correlated impurity systems in Ref.~[\onlinecite{IvadyPRB2014}]~and~[\onlinecite{IvadyPRB2013}]. In the case of periodic boundary conditions and plane wave basis set, however, this approach suffers from serious limitations due the finite size-effects that cause substantial energy shifts both in the total energy and the KS eigenvalues when the fulfillment of the gKT is investigated. Furthermore, the Koopmans' theorem is valid for states of decaying density (for $r \rightarrow \infty$)\cite{Almbladh85}, and therefore its  application for bulk states is not well defined.

\subsection{Calculation of the $w$ potential strength}

By utilizing the above described connection between the hybrid-DFT, LDA+U, and the COHSEX approximation of the Hedin's equations\cite{Hedin}, we propose an alternative method to calculate the $w$ parameter of hybrid-DFT+V$_w$ scheme. 

Following the derivation of Jiang~\emph{et al.}\ \cite{RinkeLDAU10}, the quasi particle shift of localized states in the COHSEX approximation,  $\Delta \varepsilon_m^{\text{QP}} = \left\langle \phi_m \left| \Sigma_{\text{COHSEX}} - v^{\text{PBE}}_{xc} \right| \phi_m \right\rangle$, can be written as 
\begin{equation} \label{eq:QPshift}
\left. \Delta \varepsilon_m^{\text{QP}} \right|_{\text{PBE}} = \Delta V_m \approx \left( \frac{1}{2} - n_m \right) U_{\text{COHSEX}}\text{,}
\end{equation}
where  $U^{\text{COHSEX}} = \left\langle \phi_m \phi_m \left|  W\! \left( r, r',0\right) \right| \phi_m \phi_m \right\rangle$. As derived in the previous section, the potential shift due to the mixing of the exact  and semi-local exchange, or in other words the ``generalized Kohn-Sham particle shift'', $\Delta \varepsilon_m^{\text{gKS}} = \left\langle \phi_m \left| v^{\text{PBE0}}_{xc} - v^{\text{PBE}}_{xc} \right| \phi_m \right\rangle$, can be given as
\begin{equation} \label{eq:gKSshift}
\Delta \varepsilon_m^{\text{gKS}} = \Delta V_m^{\text{PBE0x}} \approx \left( \frac{1}{2} - n_m \right) U_{\text{PBE0}}\text{.}
\end{equation}
Assuming $U_{\text{real}} \approx U_{\text{COHSEX}}$, from Eq.~(\ref{eq:wdef}), (\ref{eq:QPshift}), and (\ref{eq:gKSshift}) we obtain
\begin{eqnarray}
& w_{\text{single}} = \frac{2}{1 - 2 n_m}\left( \left. \Delta \varepsilon_m^{\text{QP}} \right|_{\text{PBE}} -\Delta \varepsilon_m^{\text{gKS}} \right) = \\ \nonumber
&  = \frac{2 }{1 - 2 n_m}  \left\langle \phi_m \left| \Sigma_{\text{COHSEX}} - v^{\text{PBE0}}_{xc} \right| \phi_m \right\rangle = \frac{2 }{1 - 2 n_m}  \left. \Delta \varepsilon_m^{\text{QP}} \right|_{\text{PBE0}} \text{,}
\end{eqnarray}
where $ \left. \Delta \varepsilon_m^{\text{QP}} \right|_{\text{PBE0}}$ is the quasi-particle shift of the PBE0 generalized Kohn-Sham particle energies.

In bulk systems the localized atomic-like states often form narrow bands that do not hybridize with other states. In such cases, not a single state but a whole band can be corrected by the $V_w$ potential correction. A suitable $w$ can be obtained from the band averaged quasi-particle shift as
\begin{equation} \label{eq:wdet}
w_{\text{band}} = \sum_{n\mathbf{k}} \omega_{\mathbf{k}}  \frac{2 }{1 - 2 n_{n\mathbf{k}}}  \left. \Delta \varepsilon_{n\mathbf{k}}^{\text{QP}} \right|_{\text{PBE0}} \text{,}
\end{equation} 
where $\omega_{\mathbf{k}}$ is the weight of k-point $\mathbf{k}$ in the irreducible Brillouin zone.
Note that the above definition does not rely on the variation of the occupation number, thus finite size effects are significantly reduced compared to the previous scheme.

\section{Methodology}

For our first principles DFT calculations we use plane wave basis state of 420~eV cutoff energy and projector augmented wave (PAW) method\cite{PAW,Kresse99} as implemented in the Vienna Ab initio Simulation Package (VASP)\cite{VASP,VASP2}. We combine VASP's hybrid functional and LDA+U implementaion to carry out hybrid-DFT+V$_w$ calculations. For cerium we use a PAW potential that includes a [Kr] $4d^{10}$ core.  We use a $10 \times 10 \times 10 $ $\Gamma$-point centered uniform grid to sample the Brillouin zone.
In our hybrid-DFT+V$_w$ calculations, we use two widespread functionals, the regular PBE0\cite{PBE0} and the range separated HSE06\cite{HSE03,HSE06} hybrid functionals, as base hybrid functional for the method. COHSEX calculations are carried by using VASP's implementation. For convergent quasi-particle calculations, we include virtual bands up to 90~eV energy and use 170~eV cutoff for the response function calculations.

\section{Computational results}

\begin{figure}[h!]
	\includegraphics[width=0.4\columnwidth]{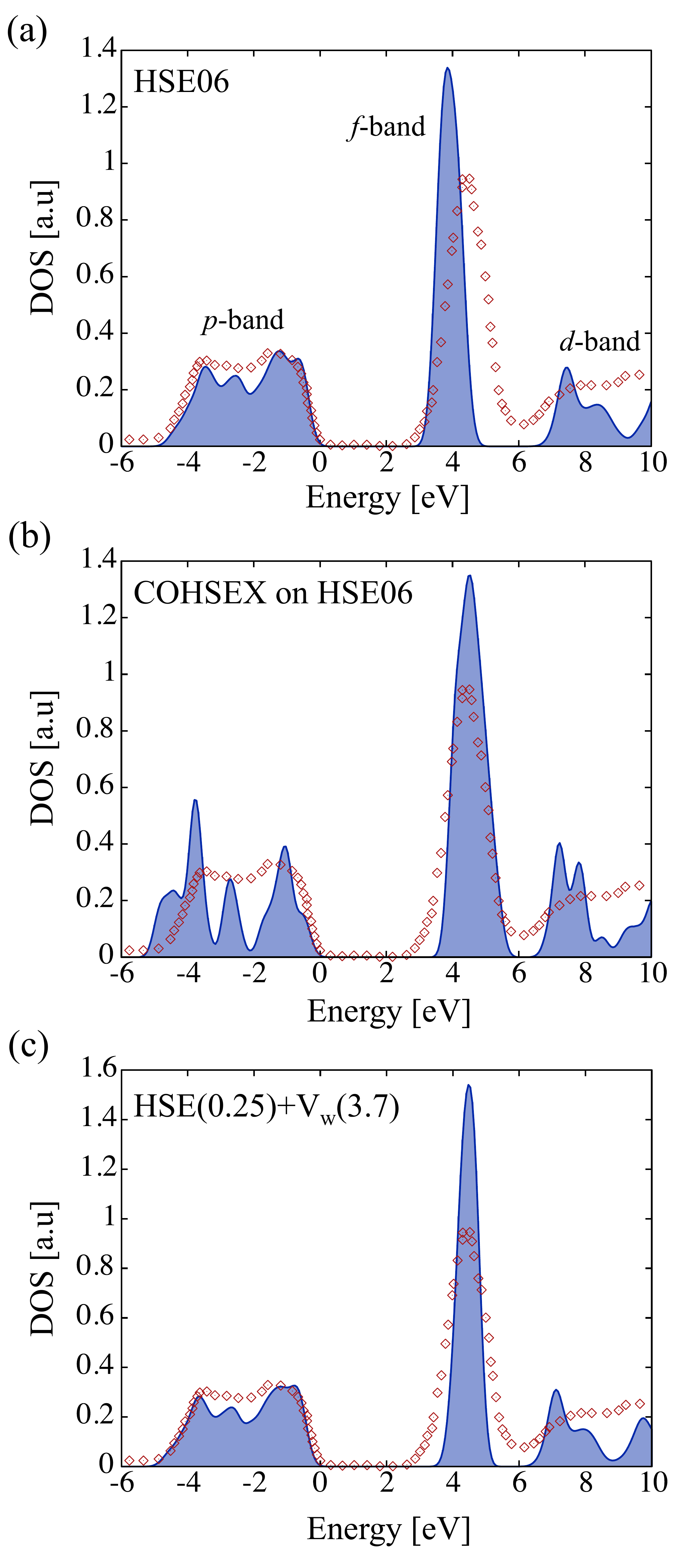}
	\caption{First principles DOS (filled curve) of cerium dioxide (CeO$_2$) as obtained by (a) HSE06 functional, (b) COHSEX approximation using HSE06 Kohn-Sham wavefunctions, and (c) HSE(0.25)+V$_w$(3.7) method. The experimental XPS-BIS spectra\cite{Wuilloud1984,Marabelli1987}  is depicted by points. The valence band edge is chosen as the origin of the energy scale in all cases. } 
	\label{fig:hse}  
\end{figure}

From the analogy of the quasi-particle equations and hybrid functional one-particle equations, one expects that hybrid-DFT+V$_w$ method can provide similar electronic structure as the COHSEX approximation when internal parameters of the former are determined properly. In the following, we investigate this expectation on the calculation of the electronic structure of CeO$_{2}$.
We focus on the electronic structure of CeO$_2$, particularly the density of the states (DOS), close to the Fermi-energy, where there are an occupied oxygen $p$-band, an unoccupied cerium $f$-band, and an unoccupied cerium $d$-band.\cite{Andersson2007, Scheffler09} The HSE06 DOS compared with the experimental XPS-BIS spectrum\cite{Wuilloud1984,Marabelli1987} is depicted in Fig.~\ref{fig:hse}(a). As can be seen in the figure, and more precisely in Table~\ref{tab:val}, HSE06 underestimates the $p$-$f$ band gap while seems to overestimate the $p$-$d$ band gap. Furthermore, it reproduces the oxygen $p$-band's band width reasonably well, however, the  Ce $f$-band is too narrow. Our findings are in agreement with previous hybrid functional studies\cite{Gillen2013}.

Starting from HSE06 generalized Kohn-Sham wave functions, the COHSEX DOS improves the $p$-$f$ and $p$-$d$ band gaps and the $f$-band width, while somewhat overestimates the $p$-band width, see Fig.~\ref{fig:hse}(b). These results suggest that using COHSEX quasi-particle shift to determine the hybrid-DFT+V$_w$ functional's internal parameters may improve on the HSE06 results. To determine the mixing parameter and the $w$ potential correction strength we apply the following procedure.

First, we adjust $w$ to correct the Ce $f$ orbitals position in the $p$-$d$ band gap. The initial value of the $w$ parameter is determined from the band averaged COHSEX quasi-particle shift of the $f$-band through Eq.~(\ref{eq:wdet}), however, additional adjustment is needed, due to the relaxation of the orbitals and the approximate nature of formula  Eq.~(\ref{eq:wdet}). Then, we vary the $\alpha$ mixing parameter of the hybrid functional to obtain the correct value of $p$-$d$ band gap.   By using  $\alpha = 0.25$ and $w = 3.7$~eV, hereinafter we name this functional as HSE(0.25)+V$_w$(3.7), we obtain an optimal fit of the DOS to  the quasi-particle DOS of CeO$_2$ calculated within COHSEX approximation. The DOS obtained by HSE(0.25)+V$_w$(3.7) functional is compared with the experimental spectra in Fig.~\ref{fig:hse}(c).  Due the adjustment of the parameters, the $p$-$d$ band gap lowers while the position of the $f$-band increases which brings the HSE(0.25)+V$_w$(3.7) results much closer to the experimental data as compared to the original HSE06 results, see Fig.~\ref{fig:hse}(c) and Table~\ref{tab:val}. Interestingly, it turns out that the mixing parameter of HSE06 should not be varied as by correcting the $f$-band's position the $d$-band lowers, due to the interaction of the bands. Similar effect has been reported in LDA+U studies\cite{Scheffler09}.  On the other hand, the HSE(0.25)+V$_w$(3.7) functional does not improve on the $f$-band's band width.  

\begin{figure}[h!]
	\includegraphics[width=0.4\columnwidth]{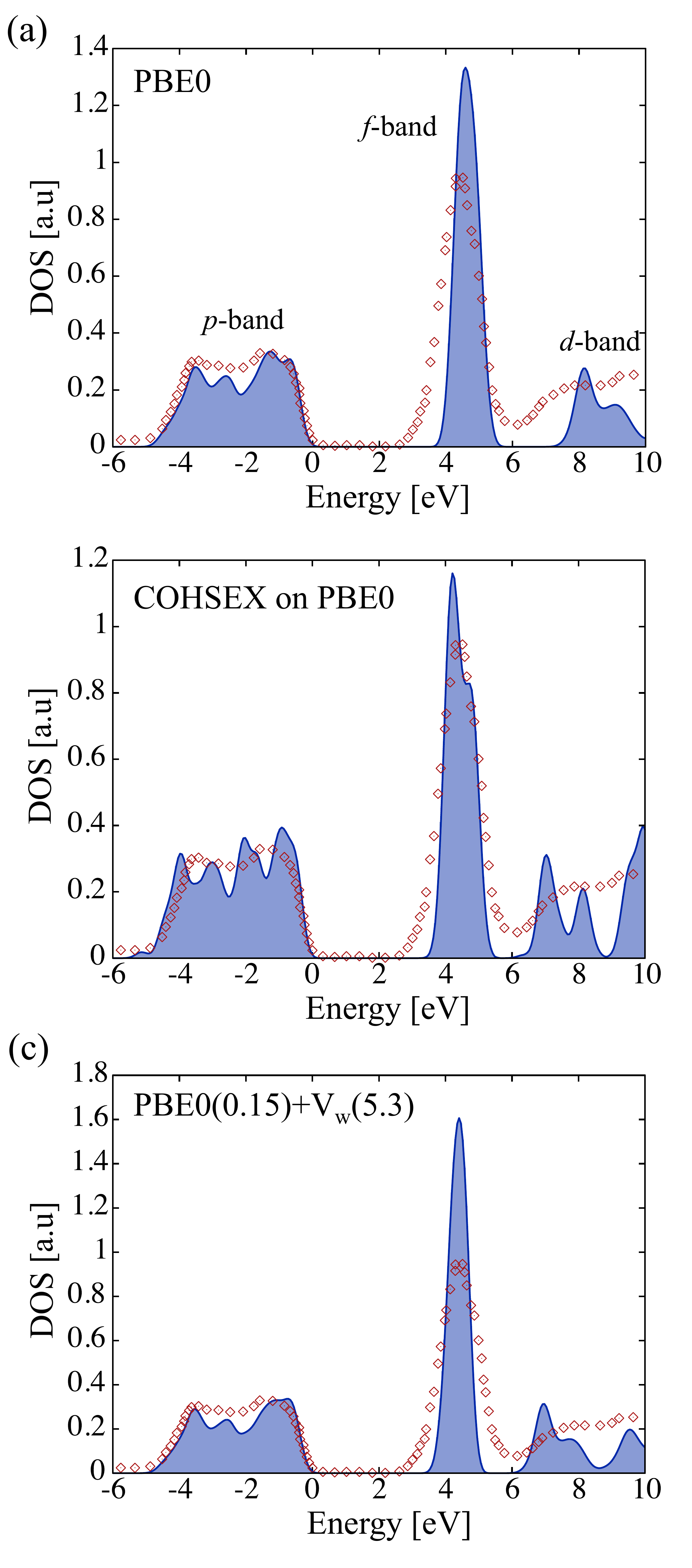}
	\caption{First principles DOS (filled curve) of cerium dioxide (CeO$_2$) as obtained by (a) PBE0 functional, (b) COHSEX approximation using PB0 Kohn-Sham wavefunctions, and (c)  PBE0(0.15)+V$_w$(5.3) method. The experimental XPS-BIS spectra\cite{Wuilloud1984,Marabelli1987}  is depicted by points. } 
	\label{fig:pbe0}  
\end{figure}

Note that in the example above the range separation parameter $\mu$ of the base HSE06 functional is not changed. To understand the effect of range separation on the DOS of CeO$_2$ in hybrid-DFT+V$_w$ method, we investigated PBE0 and PBE0 based hybrid-DFT+V$_w$ methods. In Fig.~\ref{fig:pbe0}(a), one can see the PBE0 DOS. Both the $p$-$f$ and the $p$-$d$ band gaps are overestimated, which can be due to the enhanced exchange effect introduced by the PBE0 functional. Using PBE0 orbitals, the COHSEX quasi-particle correction improves the electronic structure and the COHSEX DOS agrees well with the experimental spectra, see Fig.~\ref{fig:pbe0}(b). By using $\alpha = 0.15$ and $w = 5.3$~eV parameters in the PBE0 based hybrid-DFT+V$_w$ functional we can reproduce the quasi-particle DOS. The resultant DOS can be seen in Fig.~\ref{fig:pbe0}(c). Interestingly, PBE0(0.15)+V$_w$(5.3) DOS is nearly identical with the HSE(0.25)+V$_w$(3.7) DOS. The only difference is in position of the $f$-bands, which is slightly lower  in energy. We conclude that the range separation seems to have minor effect on the final electronic structure, however, the parameters of the hybrid-DFT+V$_w$ functional do depend on it.  

\begin{table}
\caption{ Theoretical and experimental\cite{Wuilloud1984} $p$-$f$ and $p$-$d$ band gaps in the band substructure of CeO$_2$.}
\begin{ruledtabular}
       \begin{tabular}{c|cc} 
       Method & $p$-$f$ gap & $p$-$d$ gap \\ \hline
          HSE06                              &  2.97 &  6.65 \\
          COHSEX on HSE06            &  3.48 &  6.52\\
          HSE(0.25)+V$_w$(3.7)      &  3.53 &  6.41\\ \hline
          PBE0                                  &  3.73 &  7.38 \\
          COHSEX on PBE0                & 3.43 &  6.20 \\
          PBE0(0.15)+V$_w$(5.3)     &  3.49 &  6.22 \\ \hline
          Experiment                       & 3 -- 3.5 & 5.5 -- 8
        \end{tabular}
\end{ruledtabular}\label{tab:val}
\end{table}

\section{Summary and conclusions}

We demonstrated a new approach to parameterize the orbital dependent mixing of exact exchange and semi-local exchange in the hybrid-DFT+V$_w$ scheme through the analogy of the COHSEX quasi-particel equations and the hybrid-DFT single particle equations. The computational cost of the proposed method is in the order of a hybrid-DFT functionals demand. This makes the proposed method suitable for charge self-consistent supercell calculations to study, for example, point defect. Because of the charge self-consistency and consistent functional formulation, the technique allows for the electronic structure, as well as the total energy calculations. We illustrated the performance of the method on the band structure calculation of CeO$_2$, where results achieved by the hybrid-DFT+V$_w$ method significantly improve those of either HSE06 or PBE0 functional calculations.  We note that method can be easily generalized to other semiconducting transition metal compounds.

\section*{Acknowledgments}

Support from the Knut \& Alice Wallenberg Foundation project Strong Field Physics and New States of Matter 2014-2019 (COTXS), the Grant of the Ministry of Education and Science of the Russian Federation  (grant No. 14.Y26.31.0005), and the ``Lend\"ulet program" of Hungarian Academy of Sciences is acknowledged. The calculations were performed on resources provided by the Swedish National Infrastructure for Computing (SNIC) at the National Supercomputer Centre (NSC).

\bibliographystyle{apsrev4-1}
\bibliography{references}

\end{document}